\documentclass[twocolumn,pra]{revtex4}
\usepackage[latin9]{inputenc}
\usepackage{array}
\usepackage{float}
\usepackage{bm}
\usepackage{multirow}
\usepackage{amsmath}
\usepackage{amssymb}
\usepackage{graphicx}

\makeatletter

\providecommand{\tabularnewline}{\\}

\@ifundefined{textcolor}{}
{%
 \definecolor{BLACK}{gray}{0}
 \definecolor{WHITE}{gray}{1}
 \definecolor{RED}{rgb}{1,0,0}
 \definecolor{GREEN}{rgb}{0,1,0}
 \definecolor{BLUE}{rgb}{0,0,1}
 \definecolor{CYAN}{cmyk}{1,0,0,0}
 \definecolor{MAGENTA}{cmyk}{0,1,0,0}
 \definecolor{YELLOW}{cmyk}{0,0,1,0}
}

\usepackage{dcolumn}
\usepackage{bm}
\usepackage{epsfig}
\usepackage{braket}

\makeatother

\begin{document}

\title{Selective Orientation of Chiral Molecules by Laser Fields with Twisted
Polarization}

\author{I. Tutunnikov}

\author{E. Gershnabel}

\author{S. Gold}

\author{I. Sh. Averbukh}

\affiliation{Department of Chemical and Biological Physics, Weizmann Institute
of Science, Rehovot 7610001, ISRAEL}
\begin{abstract}
We explore a pure optical method for enantioselective orientation
of chiral molecules by means of laser fields with twisted polarization.
Several field implementations are considered, including a pair of
delayed cross-polarized laser pulses, an optical centrifuge, and polarization
shaped pulses. The underlying classical orientation mechanism common
for all these fields is discussed, and its operation is demonstrated
for a range of chiral molecules of various complexity: hydrogen thioperoxide
(${\rm HSOH}$), propylene oxide (${\rm CH_{3}CHCH_{2}O}$) and ethyl
oxirane (${\rm CH_{3}CH_{2}CHCH_{2}O}$). The presented results demonstrate
generality, versatility and robustness of this optical method
for manipulating molecular enantiomers in the gas phase.
\end{abstract}
\maketitle

\section{Introduction}

\label{Introduction}

Chiral compounds are missing mirror symmetry and cannot be superimposed
with their reflection in a flat mirror \cite{UniversalChirality,Cotton,Mirror-Image}.
Chiral molecules related by such a reflection are called enantiomers.
An important research direction is chiral resolution; a problem of
differentiation of enantiomers in a mixture containing both of them.
The ability to separate the enantiomers is a crucial step in drug
synthesis, as different enantiomers of chiral drugs exhibit marked
distinctions in their biological activity \cite{ChiralDrugs}. The
related studies focus on the measurements of enantiomeric excess, handedness
of a given compound and devising techniques for manipulating mixtures
containing both enantiomers \cite{bib3,bib4,bib7,bib8,bib18,bib9,bib10,bib11,bib12,bib13,bib14,bib15}.
Traditional chiral resolution methods include crystallization, chromotography
and using enantioselective enzymes. Recently, a number of new techniques
have been developed for investigating chiral molecules in the gas
phase. This includes photoelectron circular dichroism using intense
laser pulses or synchrotron radiation \cite{bib1,bib6,bib9}, Coulomb
explosion imaging \cite{bib8,bib18,Pitzer2017}, microwave three-wave
mixing \cite{bib7,bib10,bib19,bib20}, and analysis of the laser-induced
phase shifts in the microwave signals emitted by rotating dipoles.
The later approach was theoretically suggested in \cite{bib21}, based
on quantum mechanical arguments and it relied on exciting unidirectional
rotation (UDR) of molecules with the help of a pair of nonresonant,
delayed cross-polarized laser pulses. This excitation technique had
been suggested in \cite{bib23,Kitano2009}, further experimentally
demonstrated in \cite{Korech2013,Kenta2015,JW2015}, investigated
in detail, both from the quantum and classical perspectives in \cite{Khodorkovsky2011}
and generalized to chiral trains of multiple pulses in \cite{Bloomquist2012,Valerytrain2011}.
In a recent paper \cite{Gershnabel2017} we showed that linearly polarized
laser fields whose polarization axis twists with time in some plane
are able to \textit{orient} generic asymmetric molecules. Moreover,
we found that the underlying mechanism is classical in nature. The
orientation direction is perpendicular to the plane of the polarization
rotation, and it is determined by the sense of rotation. Current femtosecond technology
offers several options for generating optical fields with twisted
polarization. A pair of delayed cross-polarized laser pulses provides
the simplest example of such a field, but there are also more complex
fields, such as optical centrifuge \cite{Corkum1999,Villeneuve2000,Mullin2011,Korobenko2014}
or polarization shaped pulses \cite{Karras2015,Prost2017}.

The electric field of the twisted light plays several roles. First of all, it induces \textit{alignment} of the most polarizable molecular
axis (for a review of laser molecular alignment, see Refs. \cite{Seidman2003,Ohshima2010,Stolow2011,Fleischer2012,Krems2013}).
 The aligned molecular axis tends to follow the rotation of the polarization vector, but due to inertia it experiences an angular lag with respect to it. The skewed rotating field induces the UDR of the axis according
to the mechanism described in \cite{bib23,Kitano2009,Khodorkovsky2011}
for linear and symmetric molecules. Moreover, as was
shown in \cite{Gershnabel2017}, in the case of generic asymmetric
molecules, the skewed twisting field also induces a torque along the
aligned molecular axis, which tends to \textit{orient} the molecules
(and their dipole moments) by rotating them about this axis. Enantiomers
of chiral asymmetric molecules are mirror images of each other \cite{UniversalChirality,Cotton,Mirror-Image}
and their physical properties, such as permanent dipole moment and
polarizability tensor inherit this symmetry relation. The components
of the molecular dipole and off-diagonal elements of the polarizability
tensor that are connected by the reflection operation have opposite
signs for different enantiomers. As a result, the above mechanical torque
along the most polarizable axis has opposite signs for different enantiomers as well \cite{Gershnabel2017}.
This leads to the counter-rotation of their permanent dipole moments, thus causing the  out-of-phase time-dependent dipole signals.
Therefore, the resulting emission from the gas bears information on
the chiral composition of the mixture \cite{bib21,Gershnabel2017}.
The classical nature of the above orientation mechanism ensures
generality and operational robustness of the related prospective techniques for detecting and separating enantiomers of chiral molecules.

In this paper, we explore the prospects of using laser fields with
twisted polarization for pure optical enantioselective orientation
of chiral molecules. Specifically, the chosen laser sources are: (i)
a pair of delayed cross-polarized laser pulses \cite{bib23,Kitano2009,Korech2013,Kenta2015,JW2015},
(ii) optical centrifuge \cite{Corkum1999,Villeneuve2000,Mullin2011,Korobenko2014}
and (iii) polarization shaped femtosecond pulses \cite{Karras2015,Prost2017}.
We examine them in application to three different chiral molecules:
hydrogen thioperoxide (${\rm HSOH}$, previously studied in \cite{bib21,Gershnabel2017}),
propylene oxide (${\rm CH_{3}CHCH_{2}O}$, the first chiral molecule
discovered in the interstellar space \cite{PPO-Science}) and a more
complex ethyl oxirane $\left({\rm CH_{3}CH_{2}CHCH_{2}O}\right)$
molecule. The structure of the paper is as following. In Sec. \ref{The Classical Model},
the studied molecules and their properties are presented, followed
by the description of the classical model of laser driven rotational
dynamics. In Sec. \ref{Results}, a comprehensive analysis of the
laser-induced enantioselective orientation is provided. Finally, Section
\ref{Conclusions} summarizes our results and discusses some potential
future developments.

\section{The Classical Model}

\label{The Classical Model} 
\begin{figure}[h]
\begin{centering}
\includegraphics[scale=0.21]{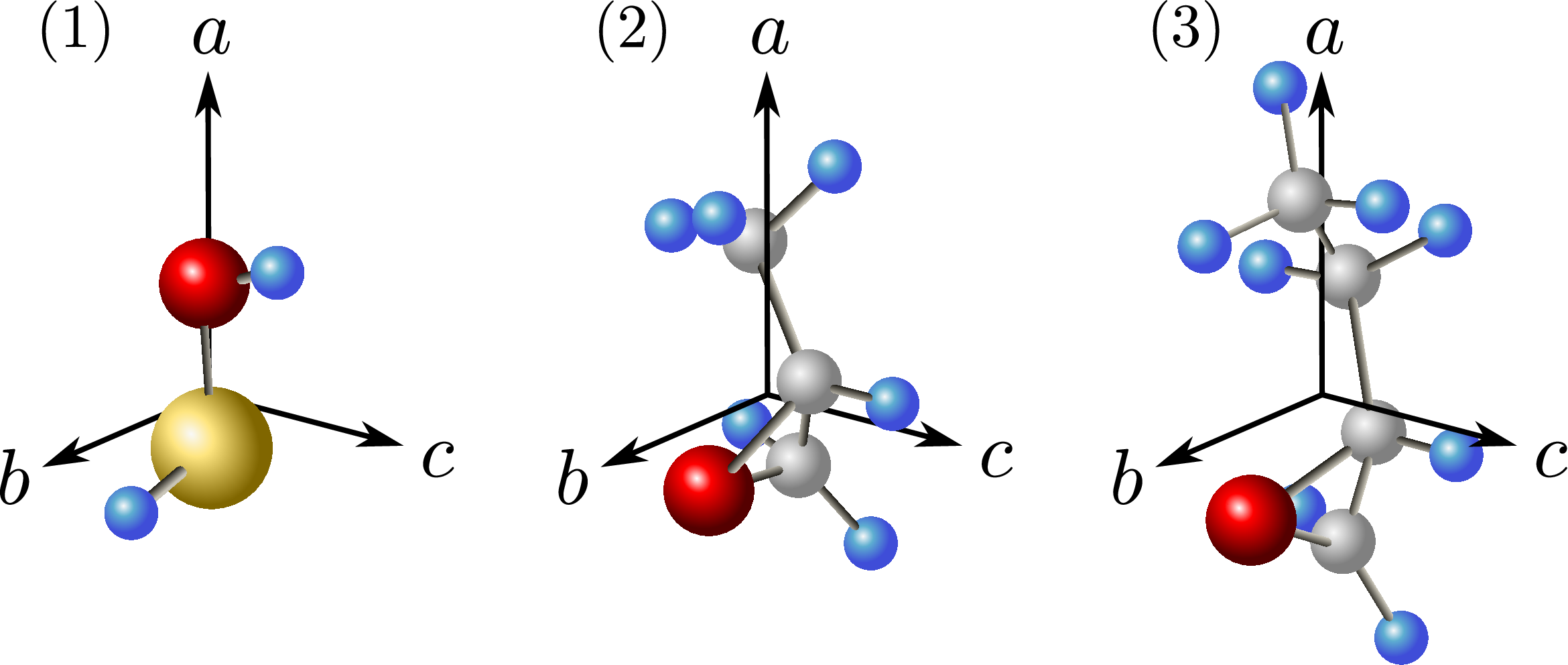}
\par\end{centering}
\caption{Molecules under consideration: (1) hydrogen thioperoxide (HSOH), (2)
propylene oxide $\left({\rm CH_{3}CHCH_{2}O}\right)$, (3) ethyl oxirane
$\left({\rm CH_{3}CH_{2}CHCH_{2}O}\right)$. Axes $a$, $b$ and $c$
are the principal axes of the molecules (the moments of inertia are ordered as $I_a<I_b<I_c$). Atoms are color-coded: gray -
carbon, blue - hydrogen, red - oxygen, yellow - sulfur.\label{fig:molecules}}
\end{figure}

In this paper, we consider chiral molecules
as asymmetric classical rigid rotors with an anisotropic polarizablity
and a permanent dipole moment, which are subject to the three excitation
schemes mentioned above. The molecules are presented in Figure \ref{fig:molecules}.

Table \ref{tab:summary-of-molecular-parameters} summarizes the molecular
properties. For computation of the molecular electronic properties we
used the GAUSSIAN software package (method: CAM-B3LYP/aug-cc-pVTZ)
\cite{Gaussian}.\\

\begin{table}[h]
\begin{centering}
\begin{tabular}{>{\centering}m{1.3cm}|>{\centering}p{1.7cm}|>{\centering}p{3.2cm}|>{\centering}p{1.9cm}}
{\scriptsize{}Molecule}  & {\scriptsize{}Moments of inertia}  & {\scriptsize{}Polarizability tensor components}  & {\scriptsize{}Dipole moment components}\tabularnewline
\hline
\multirow{3}{1.3cm}{{\scriptsize{}Hydrogen thioperoxide } } & \multirow{1}{1.7cm}{{\scriptsize{}$I_{a}=16215$} } & \multicolumn{1}{l|}{{\scriptsize{}$\alpha_{aa}=32.04$ $\alpha_{ab}=-1.21$} } & \multirow{1}{1.9cm}{{\scriptsize{}$\mu_{a}=0.020$}}\tabularnewline
 & \multirow{1}{1.7cm}{{\scriptsize{}$I_{b}=215759$ } } & \multirow{1}{3.2cm}{{\scriptsize{}$\alpha_{bb}=26.72$ $\alpha_{ac}=0.65$ } } & \multirow{1}{1.9cm}{{\scriptsize{}$\mu_{b}=0.765$}}\tabularnewline
 & \multirow{1}{1.7cm}{{\scriptsize{}$I_{c}=221933$ } } & \multirow{1}{3.2cm}{{\scriptsize{}$\alpha_{cc}=26.78$ $\alpha_{bc}=-0.03$} } & \multirow{1}{1.9cm}{{\scriptsize{}$\mu_{c}=1.435$}}\tabularnewline
\hline
\multirow{3}{1.3cm}{{\scriptsize{}Propylene oxide } } & \multirow{1}{1.7cm}{{\scriptsize{}$I_{a}=180386$ } } & \multirow{1}{3.2cm}{{\scriptsize{}$\alpha_{aa}=45.63$ $\alpha_{ab}=2.56$} } & \multirow{1}{1.9cm}{{\scriptsize{}$\mu_{a}=0.965$}}\tabularnewline
 & \multirow{1}{1.7cm}{{\scriptsize{}$I_{b}=493185$} } & \multirow{1}{3.2cm}{{\scriptsize{}$\alpha_{bb}=37.96$ $\alpha_{ac}=0.85$ } } & \multirow{1}{1.9cm}{{\scriptsize{}$\mu_{b}=-1.733$}}\tabularnewline
 & \multirow{1}{1.7cm}{{\scriptsize{}$I_{c}=553513$ } } & \multirow{1}{3.2cm}{{\scriptsize{}$\alpha_{cc}=37.87$ $\alpha_{bc}=0.65$ } } & \multirow{1}{1.9cm}{{\scriptsize{}$\mu_{c}=0.489$}}\tabularnewline
\hline
\multirow{3}{1.3cm}{{\scriptsize{}Ethyl oxirane} } & \multirow{1}{1.7cm}{{\scriptsize{}$I_{a}=246079$} } & \multirow{1}{3.2cm}{{\scriptsize{}$\alpha_{aa}=61.46$ $\alpha_{ab}=2.17$ } } & \multirow{1}{1.9cm}{{\scriptsize{}$\mu_{a}=0.394$}}\tabularnewline
 & \multirow{1}{1.7cm}{{\scriptsize{}$I_{b}=1001881$ } } & \multirow{1}{3.2cm}{{\scriptsize{}$\alpha_{bb}=47.97$ $\alpha_{ac}=1.14$ } } & \multirow{1}{1.9cm}{{\scriptsize{}$\mu_{b}=-1.878$}}\tabularnewline
 & \multirow{1}{1.7cm}{{\scriptsize{}$I_{c}=1108231$} } & \multirow{1}{3.2cm}{{\scriptsize{}$\alpha_{cc}=47.77$ $\alpha_{bc}=0.70$ } } & \multirow{1}{1.9cm}{{\scriptsize{}$\mu_{c}=0.470$}}\tabularnewline
\end{tabular}
\par\end{centering}
\caption{Summary of molecular properties: eigenvalues of the moment of inertia
tensor (atomic units), components of polarizability tensor (atomic
units) and components of dipole moment (Debye) in the body-fixed frame
of molecular principal axes. For the complimentary enantiomers, the
values of $\alpha_{ac}$, $\alpha_{bc}$ and $\mu_{c}$ have the opposite
sign.\label{tab:summary-of-molecular-parameters}}
\end{table}

We investigate the behavior of an ensemble of $N\gg1$ molecules with
the help of the Monte Carlo simulation in which rotational dynamics
of each molecule is treated numerically. There is a well known problem
with the numerical treatment of rotational dynamics in terms of Euler angles,
since it leads to singular equations of motion \cite{LANDAU}. Here,
we rely on the efficient singularity-free numerical technique, where
quaternions are used to parametrize the rotation \cite{Art-of-Molecular-Simulation,quaternions,Kuipers2002}.
This solves the problem and also avoids time consuming calculations
of trigonometric functions. The orientation of a rigid body is described
by a quaternion:
\[
q=\left(q_{0},q_{1},q_{2},q_{3}\right)=\left(\cos\frac{\theta}{2},\sin\frac{\theta}{2}\mathbf{p}\right),
\]
where $\mathbf{p}$ is a unit vector defining the direction of rotation
and $\theta$ is the angle of rotation about it. The rate of change
in time of a quaternion is given by
\begin{equation}
\dot{q}=\frac{1}{2}q\Omega,\label{eq:Quaternion-Equation-of-Motion}
\end{equation}
where $\Omega=\left(0,\boldsymbol{\Omega}\right)$ is a pure quaternion
\cite{Kuipers2002,quaternions} constructed from angular velocity
of the molecule, expressed with respect to the body-fixed frame of
molecular principal axes $a$, $b$ and $c$, $\boldsymbol{\Omega}=\left(\Omega_{a},\Omega_{b},\Omega_{c}\right)$.
In Eq. \ref{eq:Quaternion-Equation-of-Motion}, quaternions multiplication
rule is implied \cite{Kuipers2002,quaternions}. According to Euler
equations \cite{LANDAU}, the rate of change of the angular velocity
expressed with respect to body-fixed frame is
\begin{equation}
\overset{\text{\text{\tiny\ensuremath{\bm{\leftrightarrow}}}}}{\mathbf{I}}\dot{\boldsymbol{\Omega}}=\left(\overset{\text{\text{\tiny\ensuremath{\bm{\leftrightarrow}}}}}{\mathbf{I}}\boldsymbol{\Omega}\right)\times\boldsymbol{\Omega}+\boldsymbol{\mathrm{T}},\label{eq:Euler-Equation}
\end{equation}
where $\overset{\text{\text{\tiny\ensuremath{\bm{\leftrightarrow}}}}}{\mathbf{I}}$
is the moment of inertia tensor and $\boldsymbol{\mathrm{T}}=\left(\mathrm{T}_{a},\mathrm{T}_{b},\mathrm{T}_{c}\right)$
is the torque, both expressed with respect to the body-fixed frame.
To model the torque due to interaction with an electric field, we
transform the electric field $\boldsymbol{\mathcal{E}}$, expressed
with respect to the laboratory frame of reference, into the body-fixed
frame. The transformation law is $E=q^{c}\mathcal{E}q$, where $E=\left(0,\mathbf{E}\right)$
and $\mathcal{E}=\left(0,\boldsymbol{\mathcal{E}}\right)$ are pure
quaternions, constructed from the electric field in the body-fixed
frame and $\boldsymbol{\mathcal{E}}$, respectively. A conjugate of
a quaternion $q$ is denoted by $q^{c}$ \cite{Kuipers2002,quaternions}.
Again, quaternions multiplication rule is used in the transformation.
The induced dipole moment in the body-fixed principal axes frame is
given by $\mathbf{D}=\overset{\text{\text{\tiny\ensuremath{\bm{\leftrightarrow}}}}}{\boldsymbol{\alpha}}\mathbf{E}$,
where $\overset{\text{\text{\tiny\ensuremath{\bm{\leftrightarrow}}}}}{\boldsymbol{\alpha}}$
is the polarizability tensor in the body-fixed frame. Then, the torque
is $\boldsymbol{\mathrm{T}}=\mathbf{D}\times\mathbf{E}.$ We integrate
the system of Eqs. \ref{eq:Quaternion-Equation-of-Motion} and
\ref{eq:Euler-Equation} using the Runge-Kutta integration method.
We renormalize the quaternions at each time step in order to preserve
the unit norm \cite{Art-of-Molecular-Simulation,Kuipers2002}.

In our simulations, the initial conditions for the ensemble of molecules are set up
via a Monte Carlo procedure. The initial orientations of the molecules
and the corresponding quaternions for an isotropic ensemble are generated
by a random uniform sampling over the space of rotations \cite{SamplingSO3}.
We assume that the molecular ensemble is initially at thermal conditions,
and that the molecular angular velocities are distributed according
to:
\begin{align}
f\left(\boldsymbol{\Omega}\right)\propto\exp\left[-\frac{\boldsymbol{\Omega}^{T}\overset{\text{\text{\tiny\ensuremath{\bm{\leftrightarrow}}}}}{\mathbf{I}}\boldsymbol{\Omega}}{2k_{B}T}\right]=\prod_{i}\exp\left[-\frac{I_{i}\Omega_{i}^{2}}{2k_{B}T}\right],\label{eq:Angular-Velocity-Vector-Distribution}
\end{align}
where $i=a,b,c$. Since the kinetic energy is a scalar, it is coordinate
invariant so we express it with respect to the body-fixed frame. Here,
$T$ is the temperature of the gas and $k_{B}$ is the Boltzmann constant.

\section{Results and Discussion}

\label{Results}

Following the classical approach presented in Sec. \ref{The Classical Model},
we calculated the average $Z$-projection of the permanent dipole
moment, $\langle\mu_{Z}\rangle\left(t\right)$ as a function of time,
where $Z$ is the laboratory axis perpendicular to the $XY$-plane of the polarization twisting, and the angle brackets denote averaging over the ensemble. All the simulations
were done for samples of $N=500,000$ molecules, both at zero temperature
and at thermal conditions.

We considered the double pulse excitation scheme \cite{bib23,Kitano2009,Korech2013,Kenta2015,JW2015}
in application to the rotational dynamics of all three molecules.
More sophisticated schemes, involving optical centrifuge \cite{Corkum1999,Villeneuve2000,Mullin2011,Korobenko2014}
and polarization shaped pulses \cite{Karras2015} were examined in
application to propylene oxide and ethyl oxirane molecules.

\subsection{Enantioselective orientation by double pulse excitation scheme}

\label{ResultsDoublePulse}

\begin{figure}[h]
\begin{centering}
\includegraphics[scale=0.28]{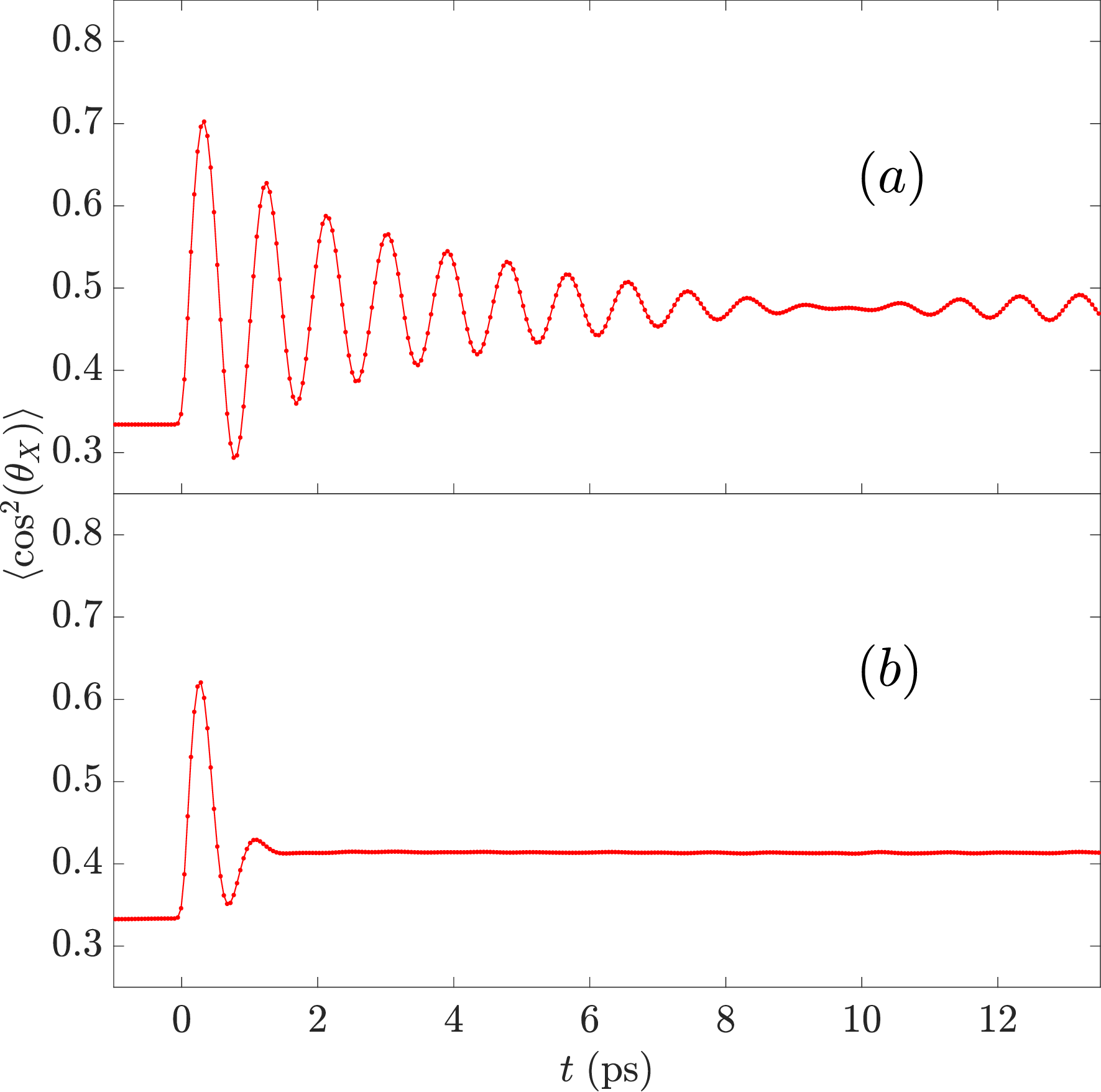}
\par\end{centering}
\caption{Alignment factor as a function of time, $\langle\cos^{2}(\theta_{X})\rangle(t)$
for hydrogen thioperoxide molecules. (a) $T=0K$, first alignment
maximum at $t=0.32\;\mathrm{ps}$; (b) $T=50K$, first alignment maximum
at $t=0.27\;\mathrm{ps}$.\label{fig:alignment-afo-time-HSOH}}
\end{figure}

In this excitation scheme, the molecules are subject to a pair of
delayed cross-polarized laser pulses. The pulses are Gaussian-shaped in time.
The peak intensity of the pulses is $I_{0}=17.7\times10^{12}\ {\rm W/cm}^{2}$,
their duration (FWHM)  is $0.1\ {\rm ps}$, which is short
compared with the typical periods of the molecular rotation. Both pulses
 propagate along the $Z$ axis. The first pulse has its maximum at $t=0$, and is polarized
along the $X$ direction, while the polarization vector of the second
pulse is in the $XY$ plane, at $45^{\circ}$ to the $X$ direction.
The first pulse induces alignment of the most polarizable axis of
the molecule along the $X$ direction \cite{Seidman2003,Ohshima2010,Stolow2011,Fleischer2012,Krems2013}.
Figure \ref{fig:alignment-afo-time-HSOH} shows the time dependent
alignment factor, $\langle\cos^{2}(\theta_{X})\rangle(t)$ for hydrogen
thioperoxide molecules. Here, $\theta_{X}$ is the angle between the
most polarizable \textit{a}-axis of the molecules and the polarization
direction of the first pulse ($X$ axis). At zero initial temperature
(Fig. \ref{fig:alignment-afo-time-HSOH}a), the alignment factor demonstrates
decaying oscillations with several local maxima. The first (most intense)
peak appears at $t=0.32\;\mathrm{ps}$. At $T=50K$ (Fig. \ref{fig:alignment-afo-time-HSOH}b)
a pronounced single alignment maximum appears shortly after the first
pulse, at $t=0.27\;\mathrm{ps}$. The oscillations decay rapidly due
to the thermal dispersion of rotational velocities. Notice that a
permanent alignment above the isotropic value of 1/3 is seen in the
long-term run.

The second laser pulse applied to the aligned molecules induces UDR
of the most polarizable molecular axis about the $Z$ direction. The
sense of this unidirectional motion is the same for both enantiomers.
However, since the studied molecules are asymmetric, this pulse also
initiates a rotation about the aligned molecular axis \cite{Gershnabel2017}.
The later rotation leads to the enantioselective orientation of the
permanent dipole moments perpendicular to the plane of the pulses,
along $Z$ axis \cite{bib21,Gershnabel2017}. Those effects are most
emphasized when the second pulse is applied at the instance of a well
developed alignment. In our simulations, we chose the delay of the
second pulse at the moment of the most intense peak of the alignment
factor, $\langle\cos^{2}(\theta_{X})\rangle(t)$.

\begin{figure}[h]
\begin{centering}
\includegraphics[scale=0.28]{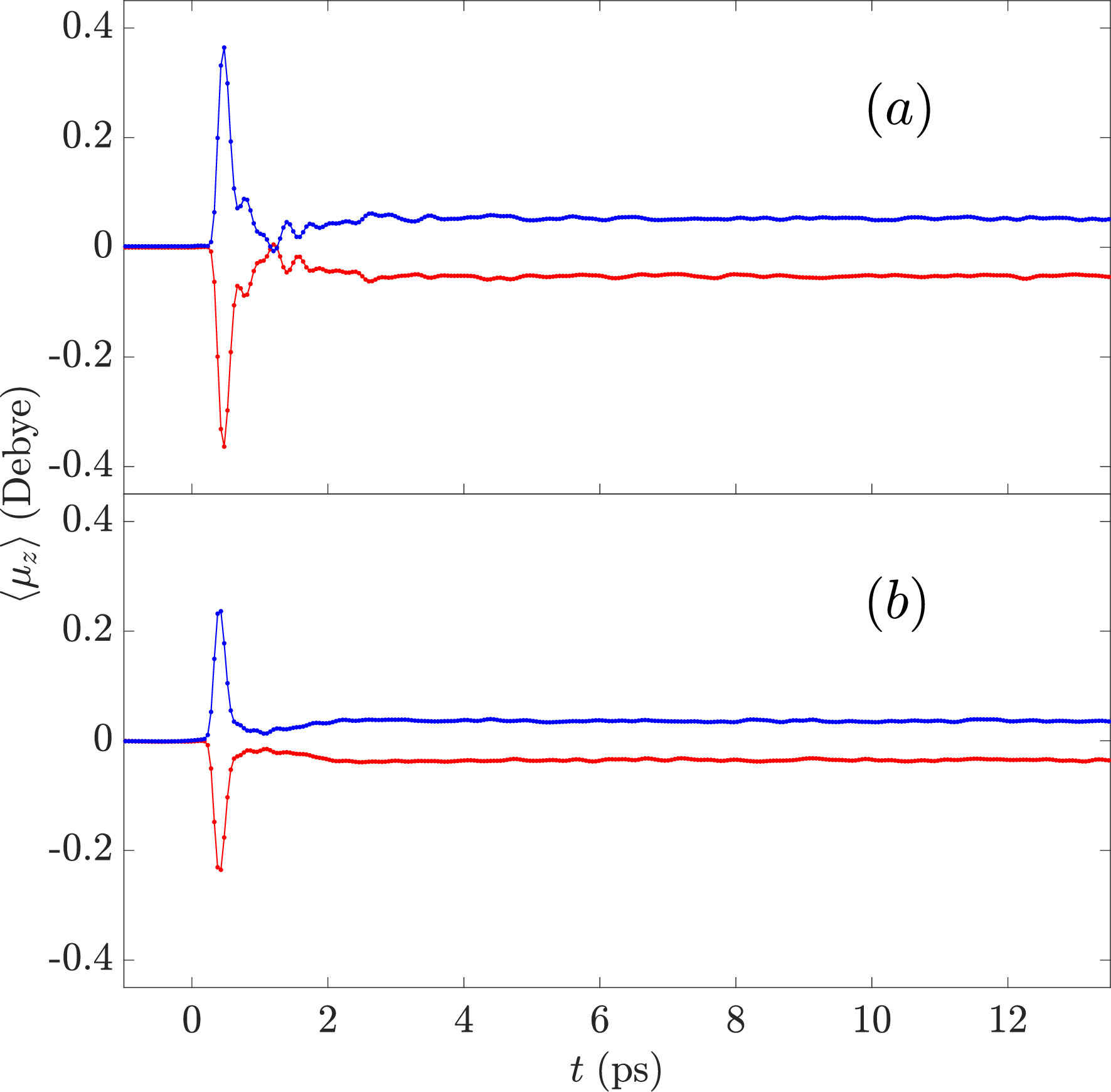}
\par\end{centering}
\caption{Average $Z$-projection of the dipole moment of both enantiomers (blue/red)
of hydrogen thioperoxide molecules as a function of time. Excitation:
double pulse scheme. (a) $T=0K$ (the delay of the second pulse is
$t=0.32\;\mathrm{ps}$). (b) $T=50K$ (the delay of the second pulse
is $t=0.27\;\mathrm{ps}$).\label{fig:HSOH-double-pulse-combined}}
\end{figure}

Figure \ref{fig:HSOH-double-pulse-combined}a shows the ensemble-averaged
dipole moment, $\langle\mu_{Z}\rangle(t)$ for two enantiomers of
hydrogen thioperoxide molecule, subject to the double pulse excitation.
As seen in the figure, $\langle\mu_{Z}\rangle$ is zero just after
the first pulse, which is consistent with the symmetry of the pulse
interaction with the induced molecular polarization. However, a pronounced dipole
signal emerges shortly after the application of the second pulse.
Moreover, one can observe an out-of-phase evolution of the dipole
signals produced by different enantiomers. The reason for such a behavior
stems from the mirror symmetry of the enantiomers, namely by the fact that the components
of the molecular dipole and off-diagonal elements of the polarizability
tensor, which are connected by reflection have opposite signs for different
enantiomers \cite{bib21,Gershnabel2017}.

Figure \ref{fig:HSOH-double-pulse-combined}b shows $\langle\mu_{Z}\rangle(t)$
for hydrogen thioperoxide molecules at $T=50K$. Here, the
magnitude of the dipole signals is lower and fine features of the
curve are washed out due to the initial dispersion of angular velocities.
Figure \ref{fig:HSOH-double-pulse-combined} shows that the most intense
dipole signals appear shortly after the second pulse. However,
one can observe a permanent orientation of the dipole moment long
after the end of the second pulse. The average value of this permanent
orientation is comparable to the peak values achieved just after the
second pulse. Long-term permanent orientation at field-free conditions
does not exist for linear polar molecules, since the individual dipoles
of the molecules are always perpendicular to the conserved vectors
of angular momentum, and they perform planar rotation with dispersed
angular velocities. In the case of asymmetric molecules considered
here, they perform a precession-like free motion. As a result, a non-zero
time averaged projection of the dipole moment along the oriented
angular momentum exists, leading to the observed permanent orientation.

\begin{figure}[h]
\begin{centering}
\includegraphics[scale=0.28]{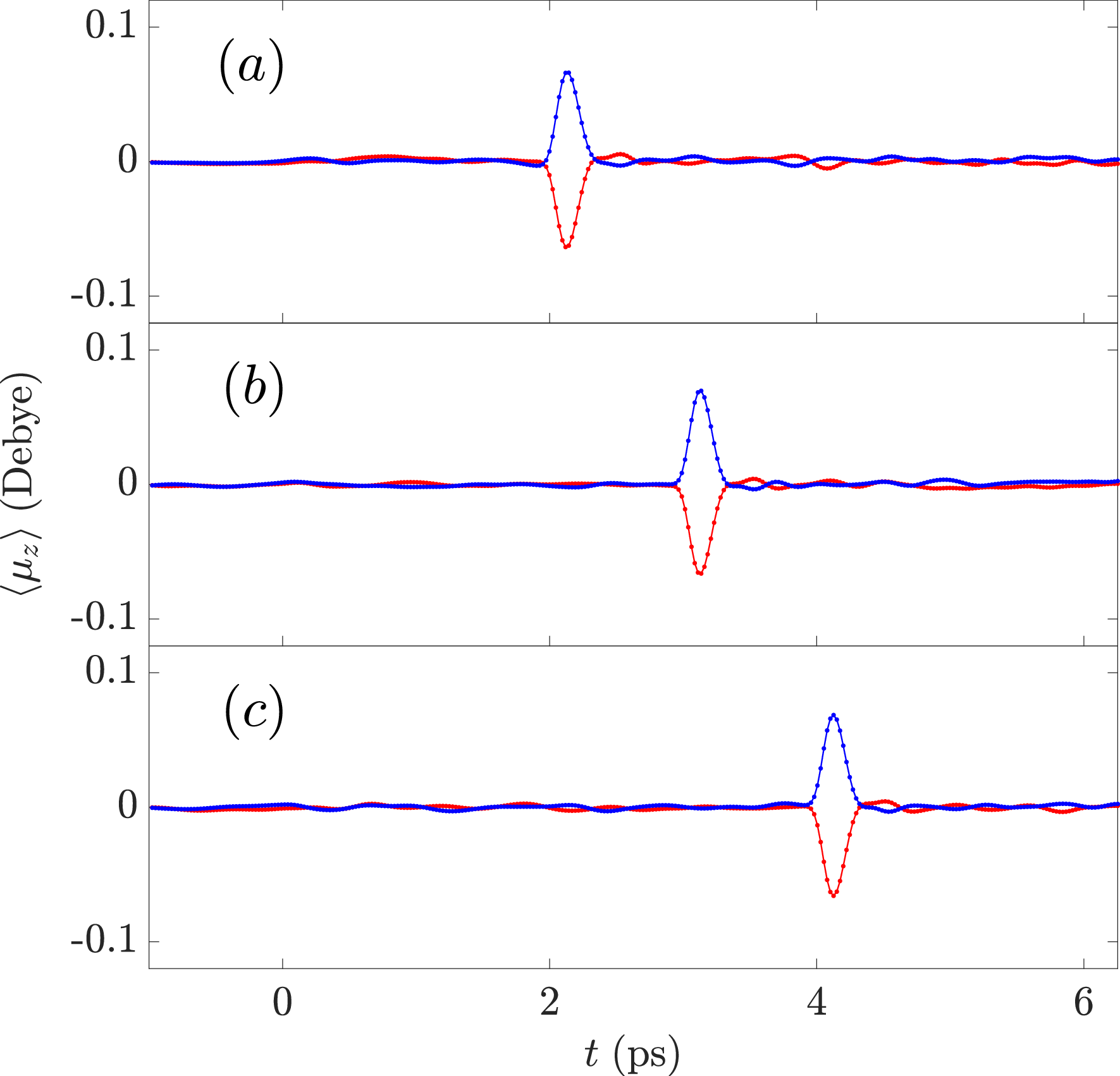}
\par\end{centering}
\caption{Average $Z$-projection of the dipole moment of both enantiomers (blue/red)
of hydrogen thioperoxide molecules as a function of time, $T=50K$.
Excitation: double pulse scheme. Delay of the second pulse is: (a)
$t=2\;\mathrm{ps}$, (b) $t=3\;\mathrm{ps}$, (c) $t=4\;\mathrm{ps}$.
\label{fig:hot-HSOH-double-pulse-234}}
\end{figure}

Figure \ref{fig:alignment-afo-time-HSOH}b shows that even a single
laser pulse creates a permanent field-free alignment (not orientation)
in the molecular ensemble. This permanent alignment may serve as a
resource for orienting molecules with the help of the second laser
pulse via the same mechanism as described above. Clearly, in this
regime the exact timing of the second pulse is not important, as illustrated
in Figure \ref{fig:hot-HSOH-double-pulse-234}. Notice that here the
permanent dipole moment is reduced compared to Figure \ref{fig:HSOH-double-pulse-combined},
since the delay of the second pulse in all three shown cases is far from the
optimal ($t=0.27\;\mathrm{ps}$).

\begin{figure}[h]
\begin{centering}
\includegraphics[scale=0.28]{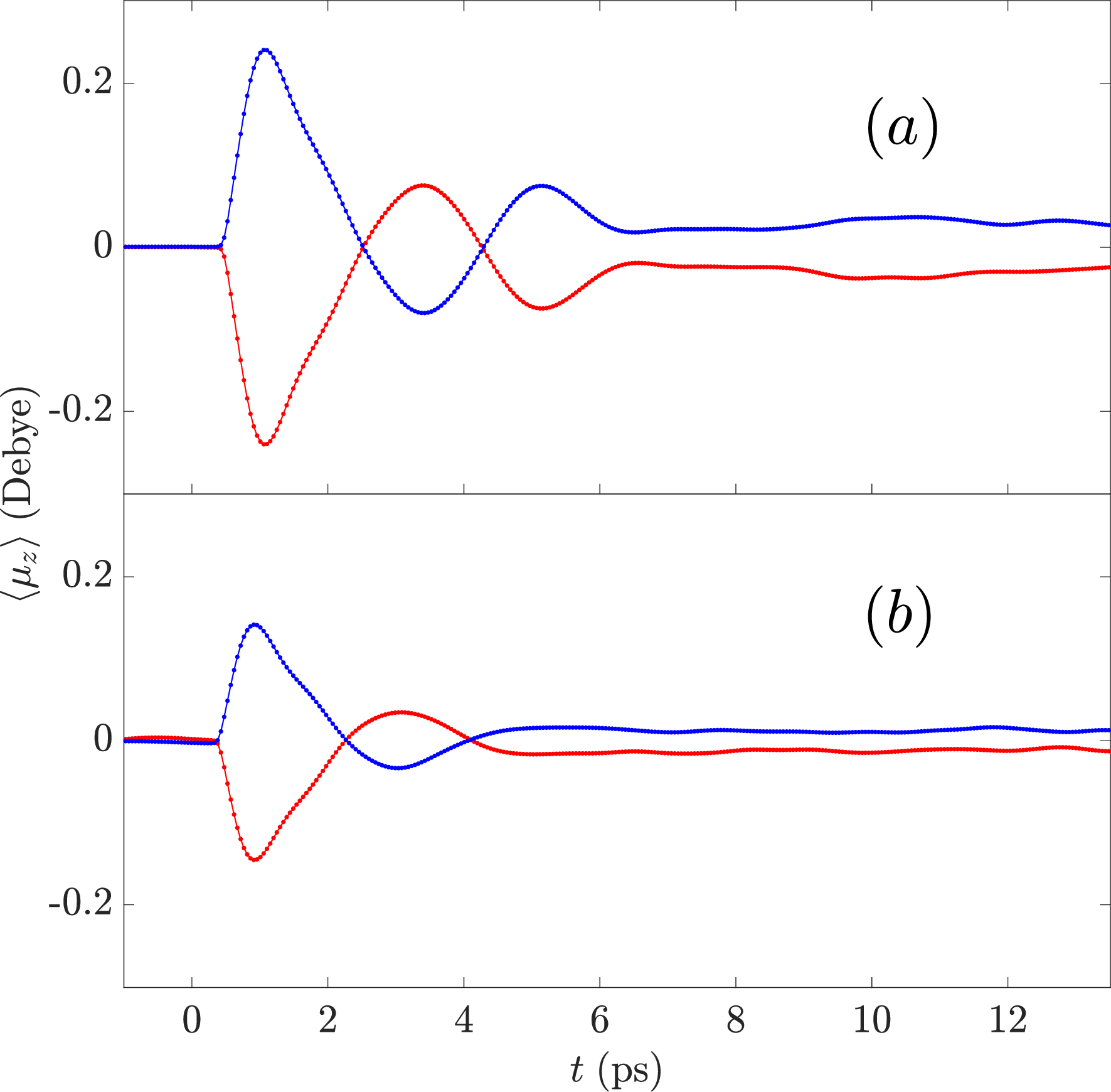}
\par\end{centering}
\caption{Average $Z$-projection of the dipole moment of both enantiomers (blue/red)
of propylene oxide molecules as a function of time. Excitation: double
pulse scheme. (a) $T=0K$ (the delay of the second pulse is $t=0.45\;\mathrm{ps}$).
(b) $T=50K$ (the delay of the second pulse is $t=0.38\;\mathrm{ps}$).\label{fig:PPO-double-pulse-combined}}
\end{figure}

\begin{figure}[h]
\begin{centering}
\includegraphics[scale=0.28]{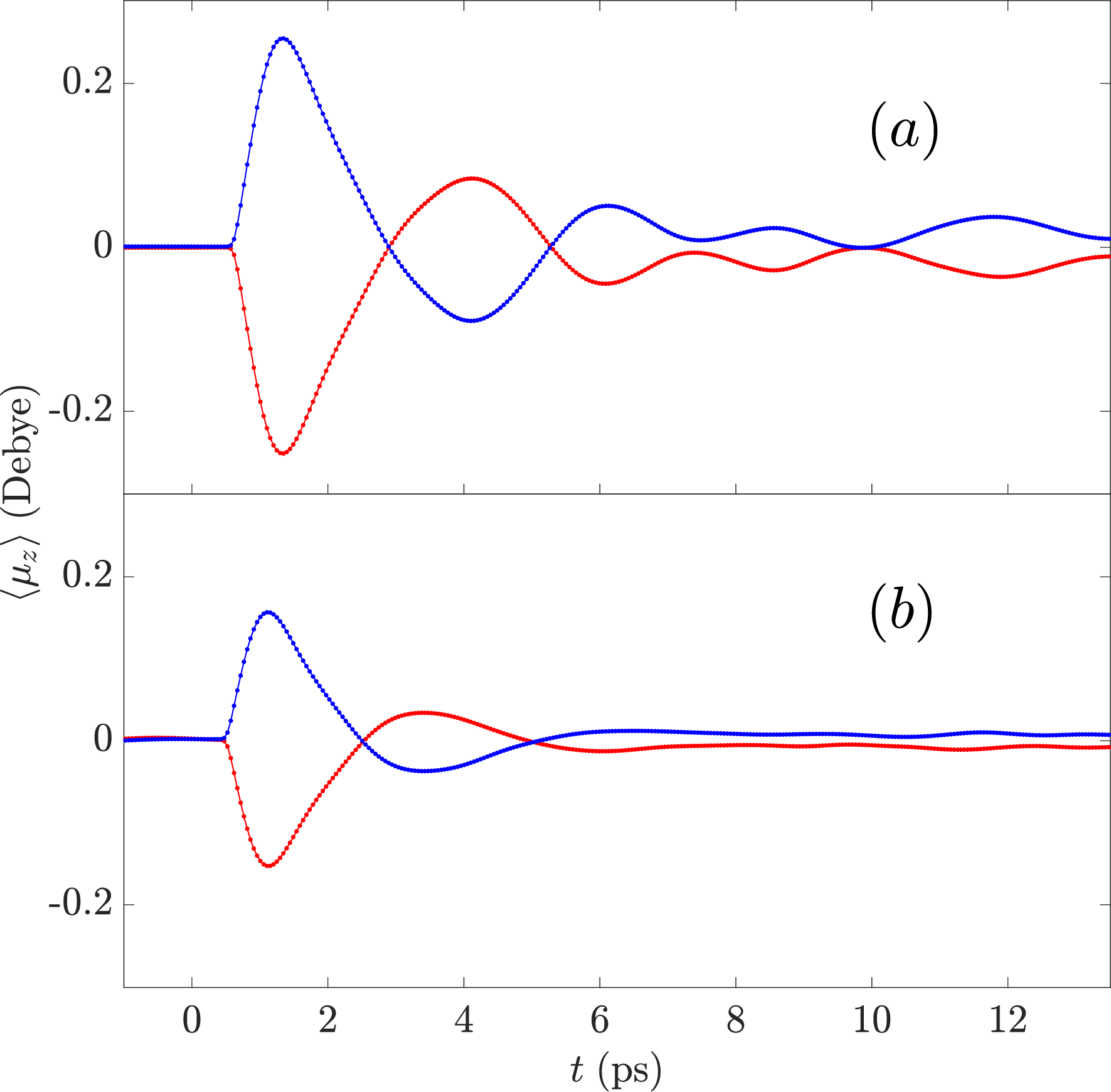}
\par\end{centering}
\caption{Average $Z$-projection of the dipole moment of both enantiomers of
ethyl oxirane molecules as a function of time. Excitation: double pulse scheme.
(a) $T=0K$ (the delay of the second pulse is $t=0.60\;\mathrm{ps}$).
(b) $T=50K$ (the delay of the second pulse is $t=0.50\;\mathrm{ps}$).\label{fig:EOX-double-pulse-combined}}
\end{figure}

We performed a similar analysis for the two additional molecules -
propylene oxide and ethyl oxirane excited by a pair of cross-polarized
pulses. Table \ref{tab:times-of-second-maxima-in-alignment} shows
the times of the first maximum of the alignment factor for all three
molecules considered in this paper. Figure \ref{fig:PPO-double-pulse-combined}
shows $\langle\mu_{Z}\rangle(t)$ for both enantiomers of the propylene
oxide molecules at $T=0K$ and $T=50K$. Similar results for ethyl
oxirane molecules are presented in Figure \ref{fig:EOX-double-pulse-combined}.
The qualitative behavior of the graphs is similar to that observed
in the case of hydrogen thioperoxide molecules, although the characteristic
timescales are longer, because of the higher moments of inertia.

\begin{table}[h]
\begin{centering}
\begin{tabular}{c|c|c}
Molecule  & $T=0K$  & $T=50K$\tabularnewline
\hline
Hydrogen thioperoxide  & $0.32$  & $0.27$\tabularnewline
Propylene oxide  & $0.45$  & $0.38$\tabularnewline
Ethyl oxirane  & $0.60$  & $0.50$\tabularnewline
\end{tabular}
\par\end{centering}
\caption{Times of first maximum in the alignment factors for the three considered
molecules. Time is measured in picoseconds and counted from the moment
of the first pulse. \label{tab:times-of-second-maxima-in-alignment}}
\end{table}

\subsection{Enantioselective orientation by optical centrifuge}

\label{ResultsOpticalCentrifuge}

\begin{figure}[h]
\begin{centering}
\includegraphics[scale=0.27]{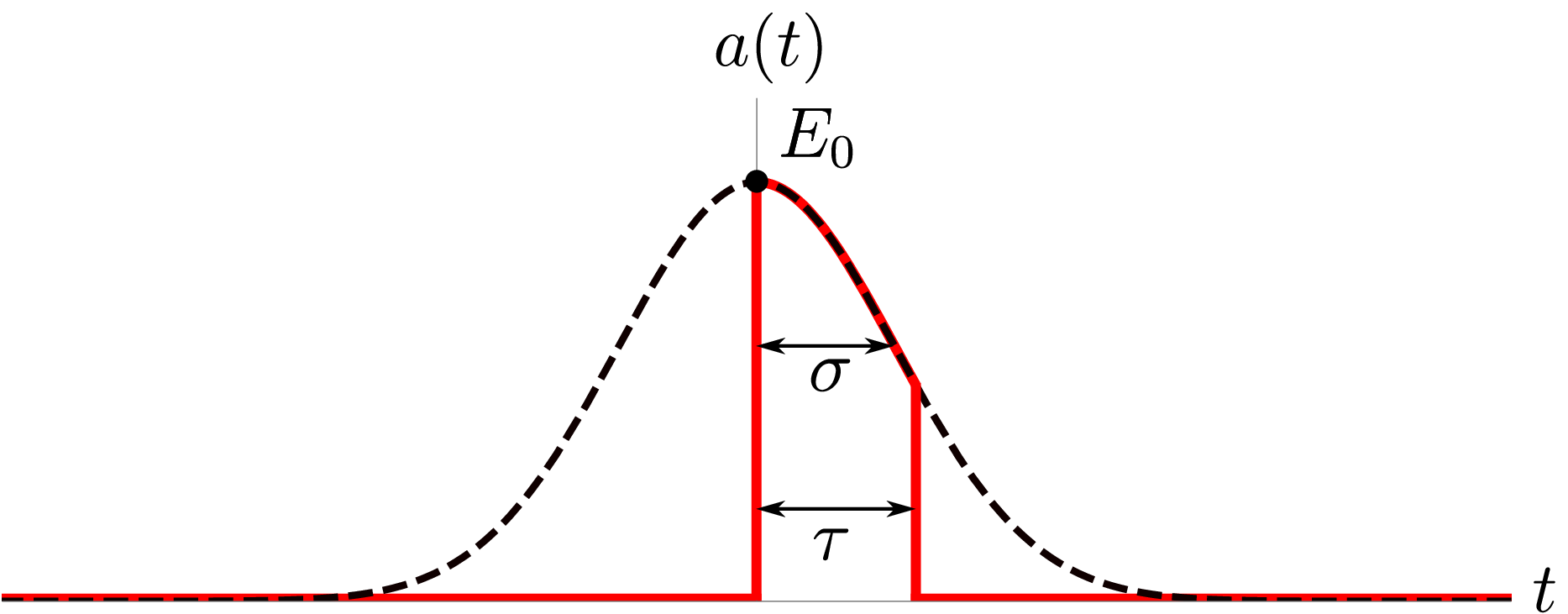}
\par\end{centering}
\caption{In red - amplitude, $a(t)$ of electric field of the centrifuge, see
equation \ref{eq:amplitude-of-centrifuge-field}. Dashed black - Gaussian
shaped envelope.\label{fig:centrifuge-amplitude}}
\end{figure}

Optical centrifuge (OC) is an optical field in which the polarization
vector rotates in plane with an acceleration, $\beta$ \cite{Corkum1999,Villeneuve2000,Mullin2011,Korobenko2014}.
Here, the field propagates along $Z$ direction while its polarization
vector rotates in the $XY$ plane. The most polarizable molecular
axis is expected to follow the accelerated rotation of the field in
the $XY$ plane. The molecules may also execute some small amplitude
out-of-plane oscillations while following the field. The question
whether a particular molecule is indeed following the polarization
vector of the field, or in other words is ``captured'' by the centrifuge,
is not a trivial one. The answer depends on the parameters of the
OC, and the initial conditions of the molecules - angular velocity
of the molecule and its initial orientation with respect to the polarization.
In the special cases of linear and symmetric molecules, the OC operation was studied
numerically in \cite{Ivanov1,Ivanov2}. Recently, an analytical study
of the OC using the theory of classical autoresonance \cite{Armon2016},
as well as its quantum mechanical counterpart \cite{Armon2017} was
done. A detailed theory for optical centrifugation of asymmetric molecules
is yet to be developed. However, it is clear at the qualitative level,
that if a molecule is indeed captured in the regime of steady angular
acceleration, its most polarizable axis follows the rotating polarization
vector with some angular lag depending on the acceleration value.
This is quite similar to the deviation of the suspended pendulum from
the vertical position when the suspension point moves with horizontal
acceleration. Thus, the situation is analogous to that of a double
pulse excitation scheme described in Subsection \ref{ResultsDoublePulse}.
The field of the optical centrifuge aligns the most polarizable axis
of the accelerating molecules close to the direction of the rotating
polarization. Because of the angular lag, the same field also induces
a torque about this direction via the mechanism described in the previous
Subsection and in \cite{Gershnabel2017}. This makes optical centrifugation
promising for enantioselective orientation. Below, we test the
feasibility of such an approach, modeling the OC with parameters of
the modern experimental setups \cite{Mullin2011,Korobenko2014}. We represent
the electric field of the optical centrifuge as
\begin{equation}
\boldsymbol{\mathcal{E}}=a(t)\left[\cos\left(2\beta t^{2}\right)\mathbf{e}_{X}+\sin\left(2\beta t^{2}\right)\mathbf{e}_{Y}\right]\cos\left(\omega t\right).\label{eq:centrifuge-field}
\end{equation}
Vectors $\mathbf{e}_{X}$ and $\mathbf{e}_{Y}$ are the basis unit
vectors in the $X$ and $Y$ directions, respectively, $\beta=0.080\;\mathrm{ps}^{-2}$
is the angular acceleration, $\omega$ is the carrier frequency of
the light. $a(t)$ is the time dependent field amplitude:
\begin{equation}
a(t)=\mathcal{E}_{0}\exp\left(-\frac{t^{2}}{2\sigma^{2}}\right)\Theta(\tau-t),\label{eq:amplitude-of-centrifuge-field}
\end{equation}
where $\mathcal{E}_{0}=0.27\times10^{10}\;\mathrm{V}/\mathrm{m}$
is the peak amplitude of the field, $\sigma=72\;\mathrm{ps}$, $\Theta$
is a unit step function, $\tau=50\;\mathrm{ps}$ is the cutoff time.
Figure \ref{fig:centrifuge-amplitude} shows the amplitude of the
electric field, $a(t)$.

The chosen parameters and the shape of the amplitude function of the
optical centrifuge are close to those used in recent experiments \cite{Milner-Private}.

\begin{figure}[H]
\begin{centering}
\includegraphics[scale=0.28]{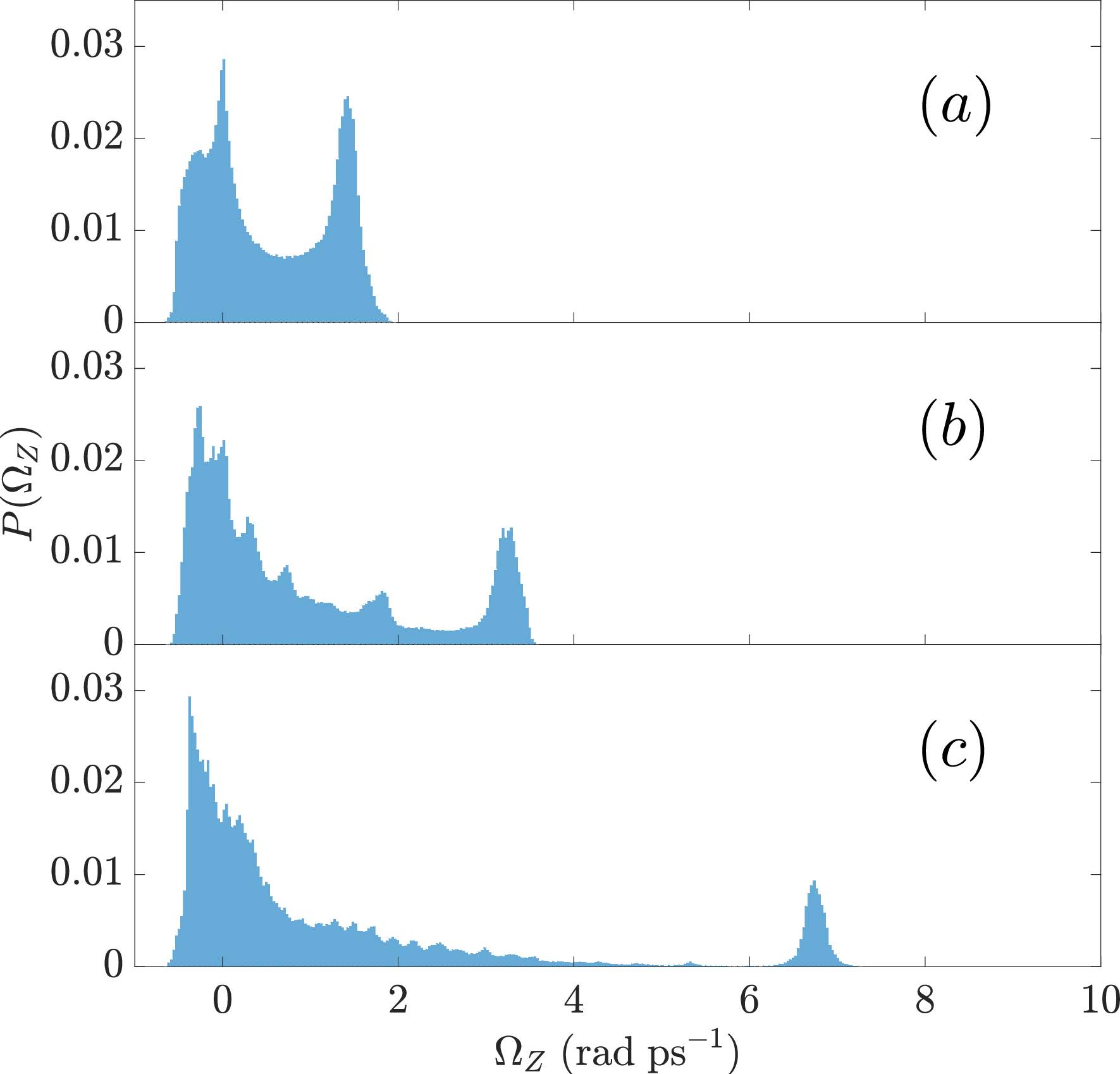}
\par\end{centering}
\caption{Projection of angular velocity of propylene oxide molecules on laboratory
$Z$ axis, $\Omega_{Z}$, at consecutive times: (a) $t=4.5\;\mathrm{ps}$,
(b) $t=10\;\mathrm{ps}$, (c) $t=20\;\mathrm{ps}$. $T=0K$. The detached
lobe corresponds to the captured molecules and constitutes about 9\%
of all the molecules (c). \label{fig:centrifuge-distributions}}
\end{figure}

\begin{figure}[h]
\begin{centering}
\includegraphics[scale=0.28]{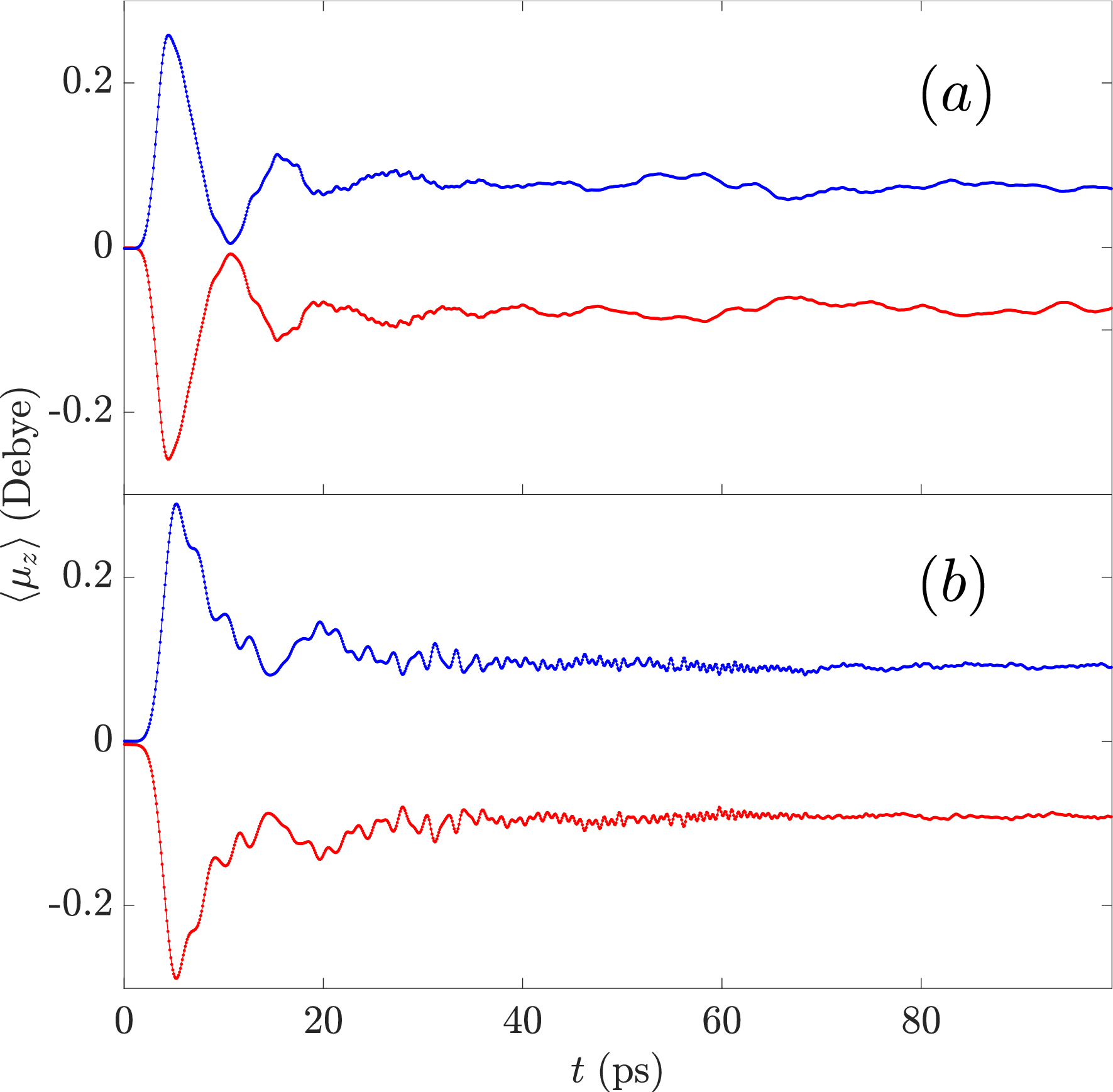}
\par\end{centering}
\caption{Average $Z$-projection of the dipole moment of both enantiomers (blue/red)
of (a) propylene oxide and (b) ethyl oxirane molecules as a function
of time. Excitation: optical centrifuge. $T=0K$. The distribution
of $\Omega_{Z}$ corresponding to the time of the peak in (a) ($t=4.5\;\mathrm{ps}$)
is shown in Figure \ref{fig:centrifuge-distributions}a.\label{fig:cold-Centrifuge-combined}}
\end{figure}

To estimate the fraction of captured molecules, we consider distribution
of the projection of angular velocity, $\Omega_{Z}$ on the laboratory
$Z$ axis. Figure \ref{fig:centrifuge-distributions} shows the distribution
of $\Omega_{Z}$ for propylene oxide molecules at consecutive times.
The initial sharp distribution spreads around zero as a result of
the gradually accelerating rotation of the electric field. As the
centrifuge speeds up, molecules throughout the distribution attempt
to follow the polarization vector and the distribution spreads even
more towards the higher values of $\Omega_{Z}$. Figure \ref{fig:cold-Centrifuge-combined}a
shows a pronounced peak in the $\braket{\mu_{Z}}(t)$ graph of propylene
oxide shortly after the initiation of the centrifuge. By the time
of the peak, the distribution of $\Omega_{Z}$ develops two separated
lobes (see Fig. \ref{fig:centrifuge-distributions}a). With time,
the lobe advances toward the higher values of $\Omega_{Z}$ and separates
completely. The lobe describes the molecules captured by the centrifuge.
In Figure \ref{fig:centrifuge-distributions}c, the separated lobe
constitutes about 9\% of all the molecules, a value typical for the
OC experiments \cite{Korobenko2014}.

The average $Z$-projection of the permanent dipole moment, $\langle\mu_{Z}\rangle(t)$
for the two enantiomers of propylene oxide are shown in Figure \ref{fig:cold-Centrifuge-combined}a.
We can observe a substantial orientation of the permanent dipole moment
in the course of the centrifuge operation. Moreover, there is a long-lasting
permanent orientation at field-free conditions after the centrifuge
cutoff time $\tau=50\;\mathrm{ps}$.

Figure \ref{fig:cold-Centrifuge-combined}b shows the average $Z$-projection
of the permanent dipole moment, $\langle\mu_{Z}\rangle$ for the two
enantiomers of ethyl oxirane molecules. Since ethyl oxirane is heavier
and has higher moments of inertia than propylene oxide, we chose a
smaller value of angular acceleration, $\beta=0.040\;\mathrm{ps}^{-2}$.
The results shown in Figure \ref{fig:cold-Centrifuge-combined}b are
qualitatively similar to those presented for propylene oxide molecule
in Figure \ref{fig:cold-Centrifuge-combined}a.

\subsection{Enantioselective orientation by polarization shaped laser pulse}

\label{ResultsPolarizaitonShaping}

\begin{figure}[h]
\begin{centering}
\includegraphics[scale=0.36]{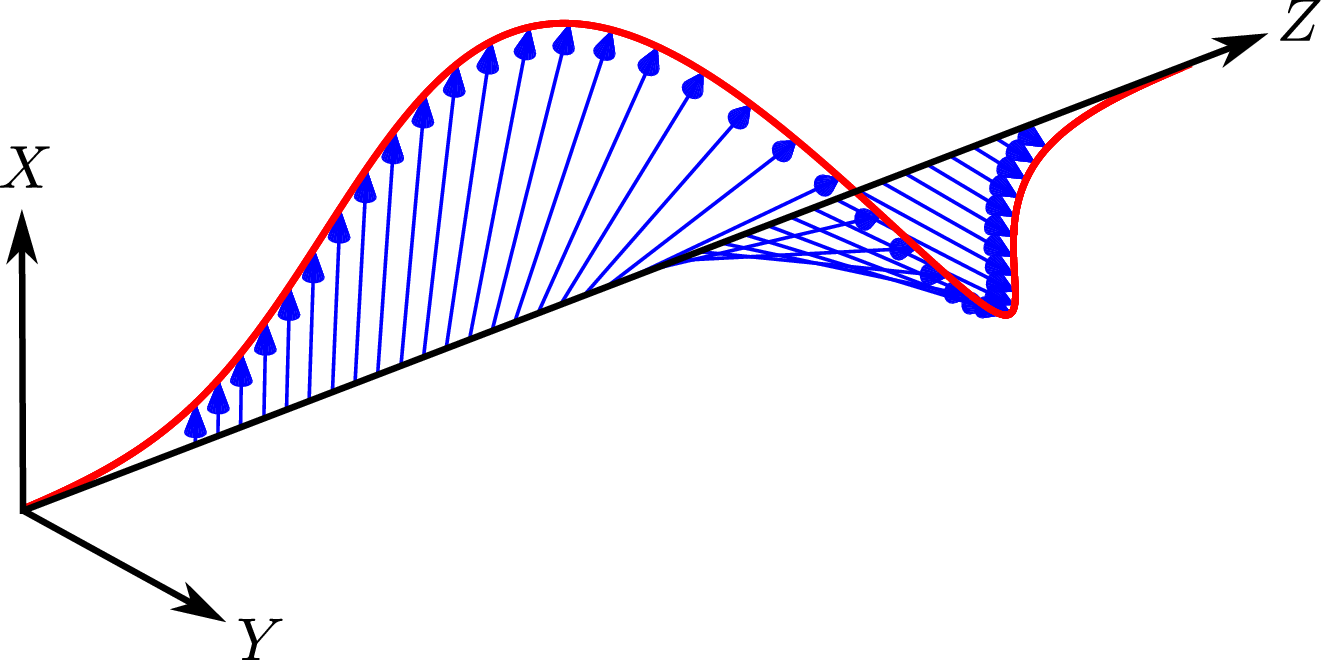}
\par\end{centering}
\caption{Illustration of polarization shaped pulse \cite{Karras2015,Prost2017}.
Here the field twists in the clockwise direction $\left(\varphi_{p}=0\right)$,
see equation $\ref{eq:polarization-shpaed-field}$.\label{fig:polarization-shaped-pulse.}}
\end{figure}

In the excitation method using polarizations-shaped pulses with continuously
twisted polarization \cite{Karras2015}, molecules are subject to
two short orthogonally polarized and partially overlapped laser pulses.
The combination of the pulses creates a field that is continuously
twisted. Here, the field propagates along $Z$ direction while its
polarization vector twists in the $XY$ plane. Initially, the polarization
vector points along the $X$ axis, and then rotates towards the $Y$
direction (see Fig. \ref{fig:polarization-shaped-pulse.}). We model
such a field as
\begin{equation}
\boldsymbol{\mathcal{E}}=\left[\mathcal{E}_{0}(t)\cos\left(\omega t\right)\mathbf{e}_{X}+\mathcal{E}_{0}(t-\tau_{p})\cos\left(\omega t+\varphi_{p}\right)\mathbf{e}_{Y}\right].\label{eq:polarization-shpaed-field}
\end{equation}
Maximum value of the Gaussian-shaped amplitude, $\mathcal{E}_{0}(t)$
is \emph{$\mathcal{E}_{0}(0)=2\times10^{8}\;\mathrm{V/cm}$}, and its
duration ($\mathrm{FWHM}$) is $0.141\;\mathrm{ps}$.
The delay between the pulses is $\tau_{p}=0.150\;\mathrm{ps}$, $\omega$
is the carrier frequency of the light. The relative phase between
the two pulses defines the sense of twisting. For $\varphi_{p}$ equal to
an integer multiple of $2\pi$, the polarization twists in the clockwise
direction, while for $\varphi_{p}$ equal to an odd multiple of $\pi$,
it twists in the counterclockwise direction.

\begin{figure}[h]
\begin{centering}
\includegraphics[scale=0.28]{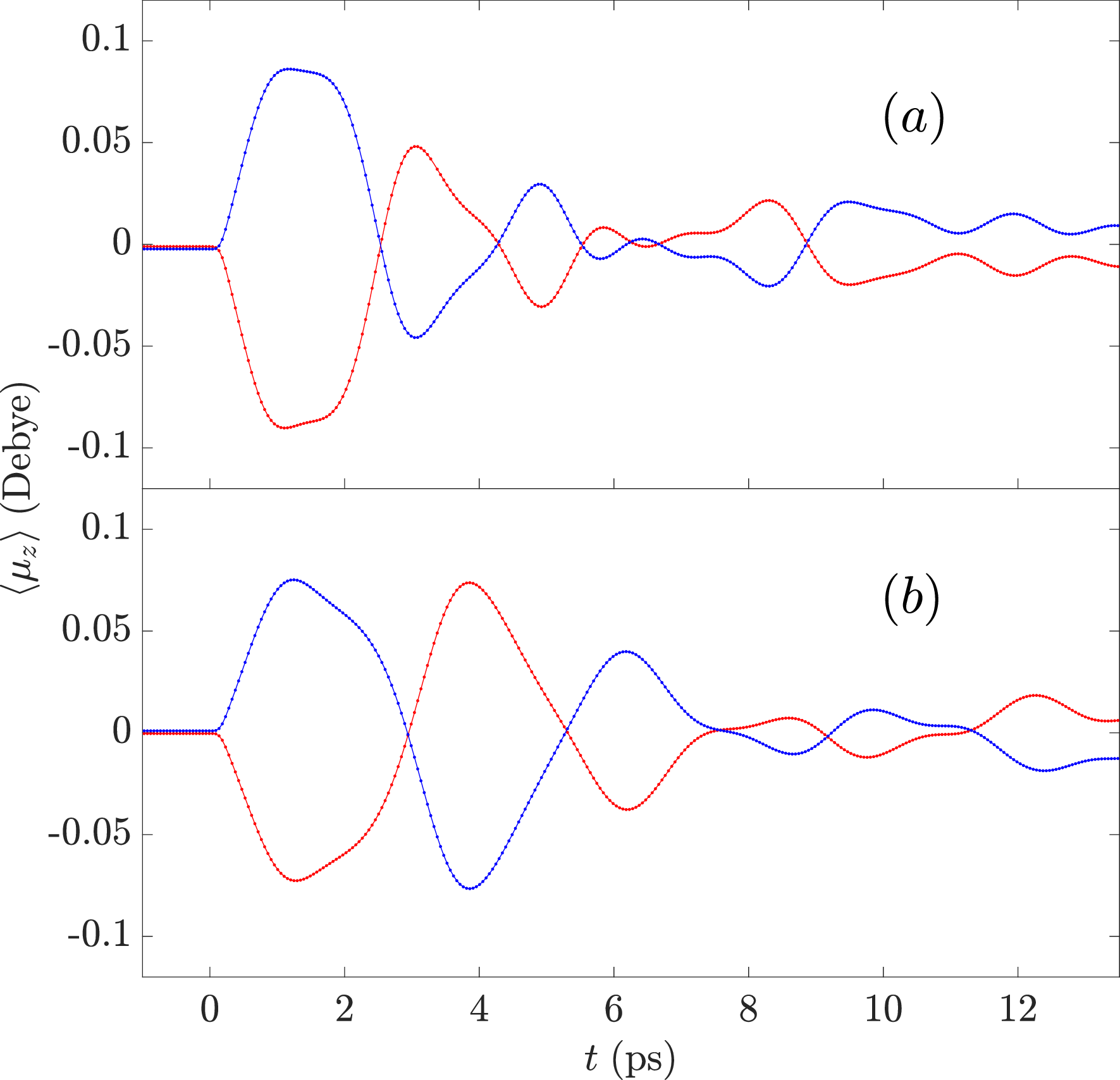}
\par\end{centering}
\caption{Average $Z$-projection of the dipole moment of both enantiomers (blue/red)
of (a) propylene oxide and (b) ethyl oxirane molecules as
a function of time. Excitation: polarization shaped pulse. $T=0K$.\label{fig:cold-French-combined}}
\end{figure}

Figure \ref{fig:cold-French-combined}a shows the time dependent average
$Z$-projection of the permanent dipole moment, $\langle\mu_{Z}\rangle(t)$,
for the two enantiomers of propylene oxide molecules. Figure \ref{fig:cold-French-combined}b
shows similar results for ethyl oxirane molecules. These dipole signals qualitatively resemble the ones already seen for two other excitation methods.

\section{Conclusions}

\label{Conclusions}

It is well known that non-resonant, linearly polarized laser pulses
may align molecules along the polarization direction, but cannot orient
them because of the symmetry of light interaction with the induced
molecular polarization.

In this paper we investigated the mechanism of orientation of generic
asymmetric molecules with laser field whose polarization twists with
time in some plane. The twisting can be either discrete in time, like
in a sequence of delayed cross-polarized laser pulses or continuous,
like in optical centrifuge or polarization shaped pulses. In all the
cases the rotating polarization induces molecular alignment along
a direction close to the polarization vector, and initiates unidirectional
rotation of the aligned molecular axis. Importantly, there is an angular
lag between the rotating polarization and the aligned molecules because
of their inertia. As it was shown in \cite{Gershnabel2017}, the skewed
field also induces rotation about the aligned molecular axis and causes
a partial orientation of the molecules and their permanent dipole
moments along the direction perpendicular to the plane of polarization
rotation. It worth mentioning that orientation of the molecular dipoles
happens here without any direct interaction between the field and
the dipoles, but it results from the controlled rotation of the molecules
to which the dipoles are attached. Moreover, this orientation is enantioselective
\cite{bib21,Gershnabel2017} and may be utilized for differentiation
of enantiomers. Our results demonstrate generality and robustness
of the enantioselective orientation mechanism utilizing various implementation
of the laser fields with twisted polarization. We examined several
specific examples of chiral molecules, including hydrogen thioperoxide,
propylene oxide and ethyl oxirane. This selective orientation can
be detected in a number of ways, ranging from direct visualization
with the help of Coulomb explosion \cite{bib8,bib18,Pitzer2017} to
observation of the out-of-phase oscillations of the enantiomers' dipole
moments \cite{bib21,Gershnabel2017}. In the latter case, the chiral
analysis is based on the measurement of laser-induced terahertz emission
from the irradiated gas samples \cite{bib7,bib10}. In addition to
the significant time-dependent dipole signals from the excited molecules,
we predict a substantial field-free permanent dipole orientation persisting
long after the laser field is over. This effect can be also used for
enantioselective measurements. The described orientation mechanism
can be optimized by tuning laser parameters, including the shape of
the laser pulses. Consideration of other types of polarization twisting
may be useful, including e.g.  chiral pulse trains \cite{Valerytrain2011} and "laser pulse enantiomers"  \cite{Steinbacher2017} (also see the references therein). The efficiency
of the methods considered here may be improved by combining polarization
twisting with other methods of molecular alignment control \cite{Seidman2003,Gershnabel2006,Gershnabel2006-2,Ohshima2010,Stolow2011,Fleischer2012,Krems2013}.
Deflecting  optically pre-oriented chiral molecules by inhomogeneous
fields  may open new ways for enantiomer separation problem.

\section{Acknowledgements}

\label{Acknowledgements}

Support by the Israel Science Foundation (Grant No. 746/15) is highly
appreciated. I.A. thanks V. Milner and T. Momose for valuable discussions
and their hospitality during his stay at the University of British
Columbia, Vancouver. We also appreciate helpful discussions with A.
Kaplan. The authors gratefully acknowledge M. Iron and P. Oulevey
for helping with the quantum-chemical calculations. I.A. acknowledges
support as the Patricia Elman Bildner Professorial Chair. This research
was made possible in part by the historic generosity of the Harold
Perlman Family.

 \bibliographystyle{unsrt}

\end{document}